\documentclass[12pt]{iopart}

\usepackage{iopams}

\usepackage{amsfonts}
\usepackage{amssymb}
\usepackage{amsthm}                                                            
\usepackage{graphicx}
\usepackage{dcolumn}
\usepackage{bm}

\begin{document}

\title {Higher-order corrections to the short-pulse equation}

\author{Levent Kurt}
\address{Department of Physical Sciences, Kingsborough Community College, \\ The City University of New York, Brooklyn, NY}
\ead{LKurt@gc.cuny.edu}

\author{Yeojin Chung}
\address{Department of Mathematics, Southern Methodist University, Dallas, TX}
\ead{ychung@smu.edu}

\author{Tobias Sch{\"a}fer}
\address{Department of Mathematics, College of Staten Island, \\ The City University of New York, Staten Island, NY}
\ead{tobias@math.csi.cuny.edu}

\begin{abstract}
        Using renormalization group techniques, we derive an extended 
        short-pulse equation as approximation to a nonlinear wave equation.
        We investigate the new equation numerically and show that the new 
        equation captures efficiently higher-order effects on pulse 
        propagation in cubic nonlinear media. We illustrate our findings using 
        one- and two-soliton solutions of the first-order short-pulse equation as
        initial conditions in the nonlinear wave equation.
\end{abstract}

\submitto{\NL}

\maketitle

\section{Introduction}

The theory of electromagnetic wave propagation in dispersive nonlinear media such as optical fibers is one of the key elements of the information age in the 20th century.  
Maxwell's equations present a complete description of the classic (non-quantized) electromagnetic field. In the presence of nonlinear effects, however, a general solution to Maxwell's equations seems impossible and approximate solutions are commonly used to explain phenomena. For the particular case of pulse propagation in cubic nonlinear media, the cubic nonlinear Schr{\"o}dinger equation (NLSE) presents a 
slowly-varying amplitude approximation that has been extremely successful in nonlinear optics \cite{agrawal:2007}. The NLSE belongs to the small class of integrable partial differential equations 
that can be solved by inverse scattering transform. It admits solitary wave solutions that show stable propagation in Maxwell's equations as long as one is working in regimes where solutions of the NLSE are close to solutions of Maxwell's equations. The extraordinary stability properties of optical solitons 
over long distances have made them promising candidates for bit carriers in fiber-optic communications \cite{essiambre-agrawal:1997}.

The derivation of the NLSE assumes a small parameter which
is the ratio of pulse width and period of the carrier wave. This small parameter is used in creating an asymptotic expansion, typically using multi-scale expansions or the renormalization group method.
In the regime of ultra-short pulses, where the pulse lengths shorten, the NLSE approximation becomes less accurate \cite{rothenberg:1992,sulem-sulem:1999}. It is possible to increase the accuracy of the NLSE approximation by
incorporating higher-order corrections. Physically, these next-order 
correction terms model important effects as Raman delay, higher-order dispersion, or self-steepening of
the pulse \cite{agrawal:2007}.

Advances in the generation of ultra-short pulses allow pulse widths in the range of 10 fs or below. The nonlinear optical pulse compression technique even makes it possible to generate pulses of about 6 fs \cite{fork-cruz-becker-shank:1987}. Femtosecond laser systems generating short pulses have led to diverse interests, studies and applications of ultra-short pulses such as extremely high intensity laser-matter interactions, characterization of high-speed electronic and optoelectronic devices and systems, optical communications, medical imaging as well as ultrafast phenomena in solid-state, chemical and biological materials \cite{weiner:1995}. 

Recently, the so-called short-pulse equation (SPE) was proposed \cite{schaefer-wayne:2004} to describe 
nonlinear pulse dynamics exploiting a scaling designed to describe the behavior of pulses in the ultra-short limit. Sakovich and Sakovich showed that the SPE is integrable \cite{sakovich-sakovich:2005}, and possesses exact one-solitary wave \cite{sakovich-sakovich:2006}, later multi-solitary wave solutions \cite{matsuno:2007} were found as well. This development led to an intensive research on the SPE over the years.
The bi-Hamiltonian structure \cite{brunelli:2005,brunelli:2006}, conserved 
quantities \cite{erbas:2008} and the periodic solutions \cite{parkes:2008} of the SPE were analyzed.
Generalizations as the vector short pulse equation \cite{pietrzyk-kanatt-bandelow:2008,sakovich:2008} and the regularized short pulse equation (RSPE) \cite{costanzino-manukian-jones:2008} were studied.
For the RSPE, the existence of multi-pulses \cite{manukian-costanzino-etal:2009}, traveling waves, and solitary wave solutions \cite{costanzino-manukian-jones:2008,costanzino:2006} were confirmed. 

It is natural to ask whether the solitary wave solutions of SPE persist in Maxwell's equations when chosen as initial conditions in
the ultra-short pulse regime, similar to the soliton solutions of NLSE in the broad pulse regime. Numerical studies show 
that this is indeed the case \cite{Levent2011}. Observed over longer distances, however, the solitary wave solutions of SPE 
show slow distortions when propagating in Maxwell's equations. This behavior is expected, as the SPE
is only an approximation to Maxwell's equations. Similar to the NLSE, it is therefore worthwhile to consider
higher-order contributions in order to obtain a better approximation to Maxwell's equation. 

The derivation
of such higher-order corrections to the SPE is the objective of the present paper. In the next section,
following the previous work \cite{chung-jones-etal:2005}, we use the renormalization group method in order
to derive the higher-order short-pulse equation (HSPE). Section \ref{sec:numerics} presents a detailed numerical study
of the new equation, comparing results to the original SPE and to Maxwell's nonlinear wave equations. 

\section{Derivation of the  higher-order SPE} \label{sec:derivation}

In this work, we analyze a one-dimensional nonlinear wave equation describing 
the propagation of pulses in nonlinear cubic media given by
\begin{equation} \label{maxwell_1d}
u_{xx}=u_{tt} + \tilde\chi_0 [u] +  \tilde\chi_3 [(u^3)_{tt}]
\end{equation}
where the $\tilde\chi_0$ and $\tilde\chi_3$ model the response of the medium to the applied field. Eq. (\ref{maxwell_1d}) is derived directly from Maxwell's equations under certain assumptions, for details we refer to previous work \cite{Levent2011}. In this work, we assume the operators $\tilde \chi_0$ and $\tilde \chi_3$ of 
particularly simple form, namely as factors in Fourier domain, meaning that
if $\hat u$ denotes the Fourier transform of $u$, 
\begin{equation} \label{cut-off}
\tilde\chi_k[\hat u] = \chi_k \hat u \cdot {\mathcal{H}}\left(|\omega|-\sqrt{\chi_0}\right), \qquad k=1,3\,.
\end{equation}
(with ${\mathcal{H}}$ being the Heaviside-function) such that unstable linear modes of (\ref{maxwell_1d}) are excluded. The numbers $\chi_k$ characterize the strength of the response of the medium.

We are interested in asymptotic solutions of the form
\begin{equation}\label{u_expansion}
u(x,t)=\epsilon A_0(\phi,x_1,x_2,...)+\epsilon^2 A_1 + ..., \qquad\phi=\frac{t-x}{\epsilon}, \qquad x_n=\epsilon^n x.
\end{equation} 
and, it has been shown \cite{schaefer-wayne:2004} that, at the leading order, the evolution equation for $A_0$ is given by
\begin{equation}\label{SPE_original}
-2 \partial_{\phi} \partial_{x_1} A_0 = \chi_0 A_0 + \chi_3 \partial^2_{\phi} A_0^3 \,.
\end{equation}
or, more precisely as
\begin{equation}\label{SPE_cut}
-2 \partial_{\phi} \partial_{x_1} A_0 = \tilde\chi_0[A_0] + \tilde\chi_3 \left[\partial^2_{\phi}A_0^3\right]
\end{equation}
where the cut-off in the operators $\tilde\chi_0$ and $\tilde \chi_3$ has been re-scaled appropriately,
hence the cut-off frequency being now $\epsilon\sqrt{\chi_0}$. In the following,
for simplicity of notation, we will write the susceptibilities as factors instead of operators.

We also note that equation (\ref{SPE_original}) can be written after a simple transform as \cite{sakovich-sakovich:2007}, 
\begin{equation}\label{spe_Transformation} 
U_{XT}=U+\frac{1}{6}U^3_{XX}\,.
\end{equation}
In the following, we compute higher-order corrections of this expansion. Although it is possible
to do this using a multi-scale expansion \cite{Levent2011}, we prefer to present the derivation using
the renormalization group (RG) method. The RG method, which was developed as a tool for asymptotic analysis \cite{chen:1994,chen:1996}, has already been successfully applied to the nonlinear wave equation (\ref{maxwell_1d}) to derive the SPE (\ref{SPE_original}) \cite{chung-jones-etal:2005} at the leading order.
 This method 
has two main advantages over the method of multiple scales: (a) algebraic calculations in the case of the RG treatment are simpler than the ones appearing in the multiple-scale expansions especially if higher-order corrections are considered, and (b) the introduction of a particular multi-scale-\textit{ansatz} is not 
required a priori for the reason that the RG equation naturally gives rise to the scales that need to be introduced for a consistent asymptotic expansion.  

We first reproduce quickly the SPE and then proceed to the next order. We apply a coordinate transformation in the form of
\begin{equation}\label{u_RGexpansion}
u(x,t) = B(\phi,x)
\end{equation}
where $\phi$ and $x$ are given in (\ref{u_expansion}). Substituting (\ref{u_RGexpansion}) in (\ref{maxwell_1d}), we obtain
\begin{equation}\label{B_1}
-\frac{2}{\epsilon} B_{\phi x} = \chi_0 B - B_{xx} + \frac{1}{\epsilon^2} \chi_3 (B^3)_{\phi \phi} \,.
\end{equation} 
We assume the expansion
\begin{equation}\label{B_1_1}
B(\phi,x) = \epsilon \Lambda_0(\phi,x) + \epsilon^2 \Lambda_1(\phi,x)  + \cdots\,.
\end{equation}
Substituting the expansion (\ref{B_1_1}) into (\ref{B_1}), the terms of $O(1)$ yield
\begin{equation}
-2(\Lambda_0)_{\phi x} = 0 \,.
\end{equation}
This implies $\Lambda_0$ is independent of $x$, i.e., $\Lambda_0=\Lambda_0(\phi)$. On the 
other hand, the terms of $O(\epsilon)$ yield 
\begin{equation}
-2(\Lambda_1)_{\phi x} = \chi_0 \Lambda_0 + \chi_3 (\Lambda_0^3)_{\phi \phi} \,.
\end{equation} 
Therefore, we find the second order approximate solution
\begin{equation}\label{sec1}
B^{(2)}(\phi,x) = \epsilon  \Lambda_0(\phi) - \frac{\epsilon^2}{2} x \left( \int_{-\infty}^{\phi} \chi_0 \, \Lambda_0(\tilde\phi)d\tilde\phi + \chi_3 \, \left(\Lambda_0(\phi)^3\right)_{\phi} \right). 
\end{equation}
Notice that a secular term appears on the equation (\ref{sec1}) which corresponds to the term proportional to $x$. In order to remove this secular term, we consider the term $\Lambda_0(\phi) - \frac{\epsilon}{2} x ( \int_{-\infty}^{\phi} \chi_0 \, \Lambda_0(\tilde\phi)d\tilde\phi + \chi_3 \, \left(\Lambda_0(\phi)^3\right)_{\phi} )$ as the Taylor expansion of a function ${\mathcal{V}}_0(\phi,x)$ about $x=0$. Thus, we need to find ${\mathcal{V}}_0(\phi,x)$ which satisfies 
%
\begin{eqnarray} 
{\mathcal{V}}_0(\phi,x=0) = \Lambda_0, \label{r1a}\\
\frac{\partial {\mathcal{V}}_0}{\partial x} = -\frac{\epsilon}{2} \left( \int_{-\infty}^{\phi} \chi_0 \, {\mathcal{V}}_0(\tilde\phi)d\tilde\phi + \chi_3 \, \left({\mathcal{V}}_0(\phi)^3\right)_{\phi} \right) \,. \label{r1b} 
\end{eqnarray}
This equation introduces a new scale $\epsilon x$. Let us define $x_1 = \epsilon x$ as in equation (\ref{u_expansion}) then we obtain
\begin{eqnarray} \label{rg1}
{\mathcal{V}}_0(\phi,x_1=0) = \Lambda_0  \label{rg1a}\\
\frac{\partial {\mathcal{V}}_0}{\partial x_1} = -\frac{1}{2} \left( \int_{-\infty}^{\phi} \chi_0 \, {\mathcal{V}}_0(\tilde\phi)d\tilde\phi + \chi_3 \, \left({\mathcal{V}}_0(\phi)^3\right)_{\phi} \right) \label{rg1b}\,. 
\end{eqnarray}
Solving equation (\ref{rg1b}) provided that the initial condition (\ref{rg1a}) is satisfied, we can express the r.h.s. of equation (\ref{sec1}) as the Taylor expansion of ${\mathcal{V}}_0$. Hence, we finally obtain the second order approximate solution, $B^{(2)}(\phi,x) = \epsilon {\mathcal{V}}_0$. Furthermore, equation (\ref{rg1b})  yields
\begin{equation}\label{nu_x1}
({\mathcal{V}}_0)_{\phi x_1} = -\frac{1}{2} \chi_0 {\mathcal{V}}_0 - \frac{1}{2} \chi_3 ({\mathcal{V}}_0^3)_{\phi\phi} \,. 
\end{equation}
This is the SPE.

For higher order corrections, we follow the similar steps. First, we need to collect higher order terms in $\epsilon$. The usual way of obtaining these is by assuming the ansatz $B = \epsilon \Lambda_0 + \epsilon^2 \Lambda_1 + \epsilon^3 \Lambda_2 + \cdots$ and collecting appropriated terms. However, this will lead to highly complicated algebraic calculations. Here, we approach this problem by assuming a different ansatz. Since we have already obtained the second order approximation of the solution, $B^{(2)}$, we assume 
\begin{eqnarray}
B(\phi,x) &=& B^{(2)} + \epsilon^{k+1} \Lambda_k(\phi,x) \,.\label{ansatz}
\end{eqnarray}
Plugging the equation (\ref{ansatz}) into equation (\ref{B_1}), we find that the next higher-order nontrivial contribution comes from the term $-B_{xx}$ in (\ref{B_1}), 
corresponding to $-\epsilon^3 ({\mathcal{V}}_0)_{x_1 x_1}$. Hence we need to choose 
$k=3$. Then, $O(\epsilon^4)$ terms yield
\begin{equation}\label{B_2}
-2(\Lambda_3)_{\phi x} = - ({\mathcal{V}}_0)_{x_1 x_1}\,.
\end{equation}
%
%
%
Before we proceed with (\ref{B_2}), let us define $N({\mathcal{V}}_0) = -({\mathcal{V}}_0)_{x_1 x_1}$ and rewrite this term.
%
From the SPE (\ref{nu_x1}), it follows that
\begin{equation}
-({\mathcal{V}}_0)_{x_1} = \frac{1}{2} \chi_0 \int_{-\infty}^{\phi} {\mathcal{V}}_0(\tilde\phi)d\tilde\phi + \frac{1}{2}\chi_3 \left({\mathcal{V}}_0(\phi)^3\right)_{\phi} \,. 
\end{equation}
Therefore, using the second derivative of ${\mathcal{V}}_0$,
\begin{equation}\label{nu_2ndDer}
-({\mathcal{V}}_0)_{x_1 x_1} = \frac{1}{2} \chi_0 \int_{-\infty}^{\phi} ({\mathcal{V}}_0)_{x_1}(\tilde{\phi})d\tilde{\phi} + \frac{1}{2}\chi_3 \left({\mathcal{V}}_0(\phi)^3\right)_{\phi x_1}, \,
\end{equation}
and the SPE (\ref{nu_x1}) along with some algebraic manipulation, we obtain $N({\mathcal{V}}_0)$,
\begin{eqnarray}
N({\mathcal{V}}_0) &=& -\frac{\chi_0^2}{4} \int_{-\infty}^{\phi}d\tilde\phi\int_{-\infty}^{\tilde\phi} {\mathcal{V}}_0(\phi^{'})d\phi^{'}   \nonumber \\
&& - \chi_0 \chi_3 {\mathcal{V}}_0^3 -\frac{3}{2} \chi_0 \chi_3 {\mathcal{V}}_0 ({\mathcal{V}}_0)_{\phi} \int_{-\infty}^{\phi} {\mathcal{V}}_0(\tilde\phi)d\tilde\phi \nonumber \\
&& - \frac{3}{2} \chi_3^2 {\mathcal{V}}_0 ({\mathcal{V}}_0)_{\phi} ({\mathcal{V}}_0^3)_{\phi} - \frac{3}{4} \chi_3^2 {\mathcal{V}}_0^2 ({\mathcal{V}}_0^3)_{\phi \phi}\,. \label{nl_op}
\end{eqnarray}
Clearly, $N({\mathcal{V}}_0)$ involves higher-order linear and nonlinear terms, a more detailed discussion will follow in the next section. Coming back to (\ref{B_2}), we obtain the 4th order approximate solution
\begin{equation}
B^{(4)}(\phi,x) = \epsilon \left( {\mathcal{V}}_0(\phi, x_1) - \frac{\epsilon^3}{2} x \int_{-\infty}^{\phi} N({\mathcal{V}}_0) d\phi^{'} \right)\,. \label{4th}
\end{equation}
In order to combine all the possible secular terms, we rewrite equation (\ref{4th}) using equations (\ref{r1a}),(\ref{r1b}) and Taylor expansion of ${\mathcal{V}}_0$. This yields
\begin{eqnarray}
B^{(4)}(\phi,x) &=&\epsilon \left( \Lambda_0(\phi) - \frac{\epsilon}{2} x \left( \int_{-\infty}^{\phi} \chi_0 \, \Lambda_0(\tilde\phi)d\tilde\phi + \chi_3 \, \left(\Lambda_0(\phi)^3\right)_{\phi} \right)
\right. \nonumber \\ && \left. - \frac{\epsilon^3}{2} x \int_{-\infty}^{\phi} N(\Lambda_0) d\phi^{'} \right) +O(\epsilon^5)\,.
\end{eqnarray}
Again, to remove the secular terms which are proportional to $x$, we now need to find $A(\phi,x)$ satisfying
\begin{eqnarray}
&&A(\phi,x=0) = \Lambda_0(\phi),\\
&&\frac{\partial A}{\partial x} = - \frac{\epsilon}{2} \left( \int_{-\infty}^{\phi} \chi_0 \, A(\tilde\phi)d\tilde\phi + \chi_3 \, \left(A(\phi)^3\right)_{\phi} \right) - \frac{\epsilon^3}{2} \int_{-\infty}^{\phi} N(A) d\phi^{'}\,.\label{rg2b}
\end{eqnarray}
The term that is proportional to $\epsilon^3$ leads us to introduce a new scale $x_3=\epsilon^3x$ in addition to $x_1 = \epsilon x$.
From equation (\ref{rg2b}), we also find
\begin{equation}
\frac{\partial A}{\partial x_1}  =  - \frac{1}{2} \left( \int_{-\infty}^{\phi} \chi_0 \, A(\tilde\phi)d\tilde\phi + \chi_3 \, \left(A(\phi)^3\right)_{\phi} \right) - \frac{\epsilon^2}{2} \int_{-\infty}^{\phi} N(A) d\phi^{'}\,.
\end{equation}
%
%
%
Denoting $\varkappa = x_1,$ we finally obtain
\begin{equation}\label{hspe}
-2A_{\varkappa \phi} = \chi_0 A+ \chi_3 (A^3)_{\phi \phi} + \epsilon^2 N(A) \,.
\end{equation}
This is the higher-order short pulse equation, and $A({\varkappa},\phi)$ is the magnitude of the electric field following the introduction of the new variable $\varkappa$. It is obvious that this equation is an extension of the SPE (\ref{SPE_original}) with additional terms on the right hand side. The additional operator $N(A)$, represents the higher-order corrections of the SPE approximation.

\section{Numerical Analysis} \label{sec:numerics}

In order to show that the HSPE captures the effects beyond the SPE approximation, we 
compare the solutions of the original nonlinear wave equation (\ref{maxwell_1d}) to the solutions of the
SPE and the HSPE. As initial conditions, we use one- and two-soliton solutions of the SPE. We first summarize very briefly the structure of these solutions 
and then discuss their evolutions in the HSPE given by (\ref{hspe}) in comparison to the nonlinear
wave equation (\ref{maxwell_1d}). Both equations are numerically solved using standard techniques.
For the nonlinear wave equation we use a pseudo-spectral leap-frog scheme in Fourier space, the 
SPE and the HSPE are solved using a fourth-order exponential time-differencing method
\cite{kassam-trefethen:2005} in Fourier space. For the solution of the nonlinear wave equation
it is essential to implement the cut-off in the susceptibility operators as in (\ref{cut-off}). For the 
HSPE, this can be done in a similar fashion. In the examples considered here, we have found only
a very little difference when neglecting the cut-off and treating the operators simply as scalars.
However, for consistency, the cut-off was implemented in all simulations.

Before we proceed, let us also note that, for the numerical simulation, it is useful to cast the HSPE in
 a slightly different form, 
\begin{equation}
A_{\varkappa} = -\frac{\chi_1}{2}\int_{-\infty}^{\phi} A(\varkappa,\tilde\phi)d\phi -\frac{\chi_3}{2}
\left(A^3\right)_{\phi} -\frac{\epsilon^2}{2}\int_{-\infty}^{\phi} N(A(\varkappa,\tilde\phi))d\tilde\phi,
\end{equation}
where we write the integral of the operator $N(A)$ involving the higher-order correction terms as
\begin{eqnarray}
-\int_{-\infty}^{\phi} N(A(\varkappa,\tilde\phi))d\tilde\phi &=& \frac{\chi_0^2}{4}\int_{-\infty}^{\phi}d\phi_1 \int_{-\infty}^{\phi_1}d\phi_2\int_{-\infty}^{\phi_2}d\phi_3
A(\varkappa,\phi_3) \nonumber \\
&&+ \frac{\chi_0\chi_3}{4} \int_{-\infty}^{\phi}d\tilde\phi A(\varkappa,\tilde\phi)^3 \nonumber \\
&&+ \frac{3\chi_0\chi_3}{4} A(\varkappa,\phi)^2\int_{-\infty}^{\phi}d\tilde\phi A(\varkappa,\tilde\phi) \nonumber \\
&&+\frac{3\chi_3^2}{4} A(\varkappa,\phi)^2\left(A(\varkappa,\phi)^3\right)_{\phi}\,.
\label{correction_terms}
\end{eqnarray}
Here, we define 
\begin{eqnarray}
L_1(A) &=& \frac{\chi_0^2}{4}\int_{-\infty}^{\phi}d\phi_1 \int_{-\infty}^{\phi_1}d\phi_2\int_{-\infty}^{\phi_2}d\phi_3
A(\varkappa,\phi_3), \\
N_1(A) &=& \frac{\chi_0\chi_3}{4} \int_{-\infty}^{\phi}d\tilde\phi A(\varkappa,\tilde\phi)^3,  \\
N_2(A) &=& \frac{3\chi_0\chi_3}{4} A(\varkappa,\phi)^2\int_{-\infty}^{\phi}d\tilde\phi A(\varkappa,\tilde\phi), \\
N_3(A) &=&  \frac{3\chi_3^2}{4} A(\varkappa,\phi)^2\left(A(\varkappa,\phi)^3\right)_{\phi}\,.
\end{eqnarray}
In the next subsection, we will compare the linear term $L_1$ to the nonlinear terms $N_1,N_2,N_3$. Also note that
numerical evaluation of these terms can be done by using the Fast Fourier Transform and that the
singularity at $\omega=0$ is removed by introducing the cut-off.

\subsection{Propagation of one-pulse solutions}

The analytical one-soliton solution of the SPE is found \cite{sakovich-sakovich:2006}
\begin{eqnarray}\label{spe_soln}  
U=4mn\frac{m\sin\psi\sinh\phi+n\cos\psi\cosh\phi}{m^2\sin^2\psi+n^2\cosh^2\phi} \\ 
X=Y+2mn\frac{m\sin2\psi - n\sinh2\phi}{m^2\sin^2\psi+n^2\cosh^2\phi}
\end{eqnarray}
with
\begin{equation}\label{spe_parameters}
\phi = m\left(Y+T\right),\quad \psi=n\left(Y-T\right), \qquad 
n = \sqrt{\left(1-m^2\right)}\,. 
\end{equation}
Here, we only consider non-singular solutions, which implies that the soliton parameter $m$ satisfies the condition $0<m<\sin\frac{\pi}{8} \approx 0.383$.

In our numerical simulations, we set $m=0.35$ in order to have strong nonlinear effects, but no singularities. We also choose $\epsilon=0.4$ since we are mainly interested in higher-order effects. Therefore,
we expect that the solutions of the nonlinear wave equation and the SPE will show differences already
after fairly short propagation distances. 

\begin{figure}[htb]
\centering
\scalebox{0.4}{\includegraphics[bb =150 50 554 570]{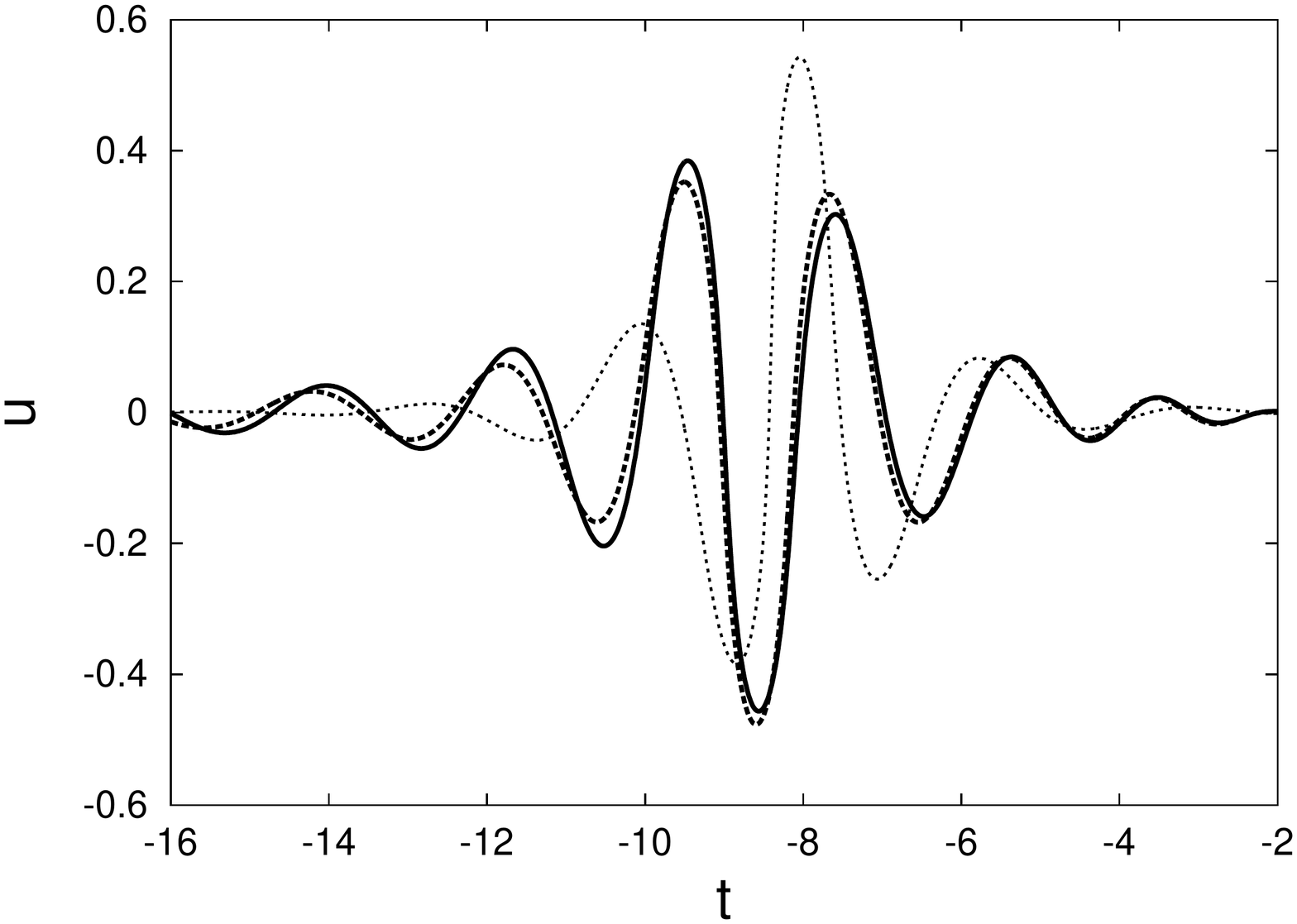}}
\hfill
\caption{Snapshot of the evolution of the SPE soliton (\ref{spe_soln}) in the nonlinear wave equation (\ref{maxwell_1d}) (solid line), the HSPE (\ref{hspe}) (dahed line) and the SPE (\ref{spe_Transformation}) (dotted line).}
\label{fig:figure1}
\end{figure}

We set a propagation distance $x=51.2$, and susceptibilities $\chi_0=2, \chi_3 = 1/3$. Using an SPE soliton (\ref{spe_soln}) as the initial condition, we numerically integrate the SPE (\ref{spe_Transformation}), the HSPE (\ref{hspe}), and the full nonlinear wave equation (\ref{maxwell_1d}).  
Figure \ref{fig:figure1} demonstrates the numerical solutions at the distance $x=51.2$ of the SPE, the HSPE, and the nonlinear wave equation. Clearly, if an SPE soliton (\ref{spe_soln}) is taken as the initial condition,
the HSPE provides a much better approximation to the nonlinear wave equation than the 
SPE.

We now consider the nonlinear operator (\ref{nl_op}) that presents the higher-order contributions. Although,
we cannot make a general statement, our numerical experiments show that for large $m$-values
close to singularity, surprisingly, the linear term seems to be dominant. 

\begin{figure}[htb]
\scalebox{0.3}{\includegraphics[bb = 0 200 654 770]{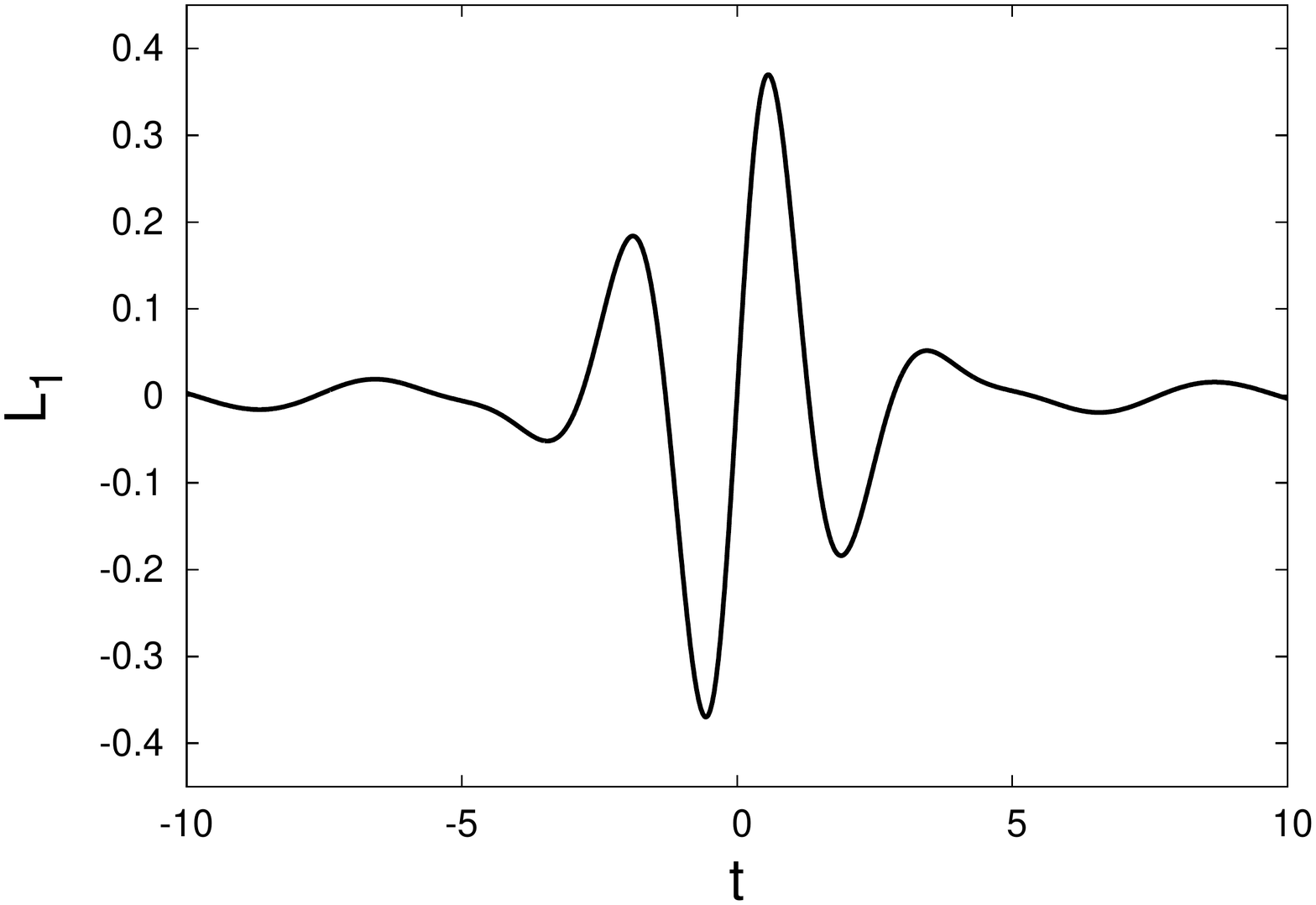}}
\hfill
\scalebox{0.3}{\includegraphics[bb = -100 200 754 770]{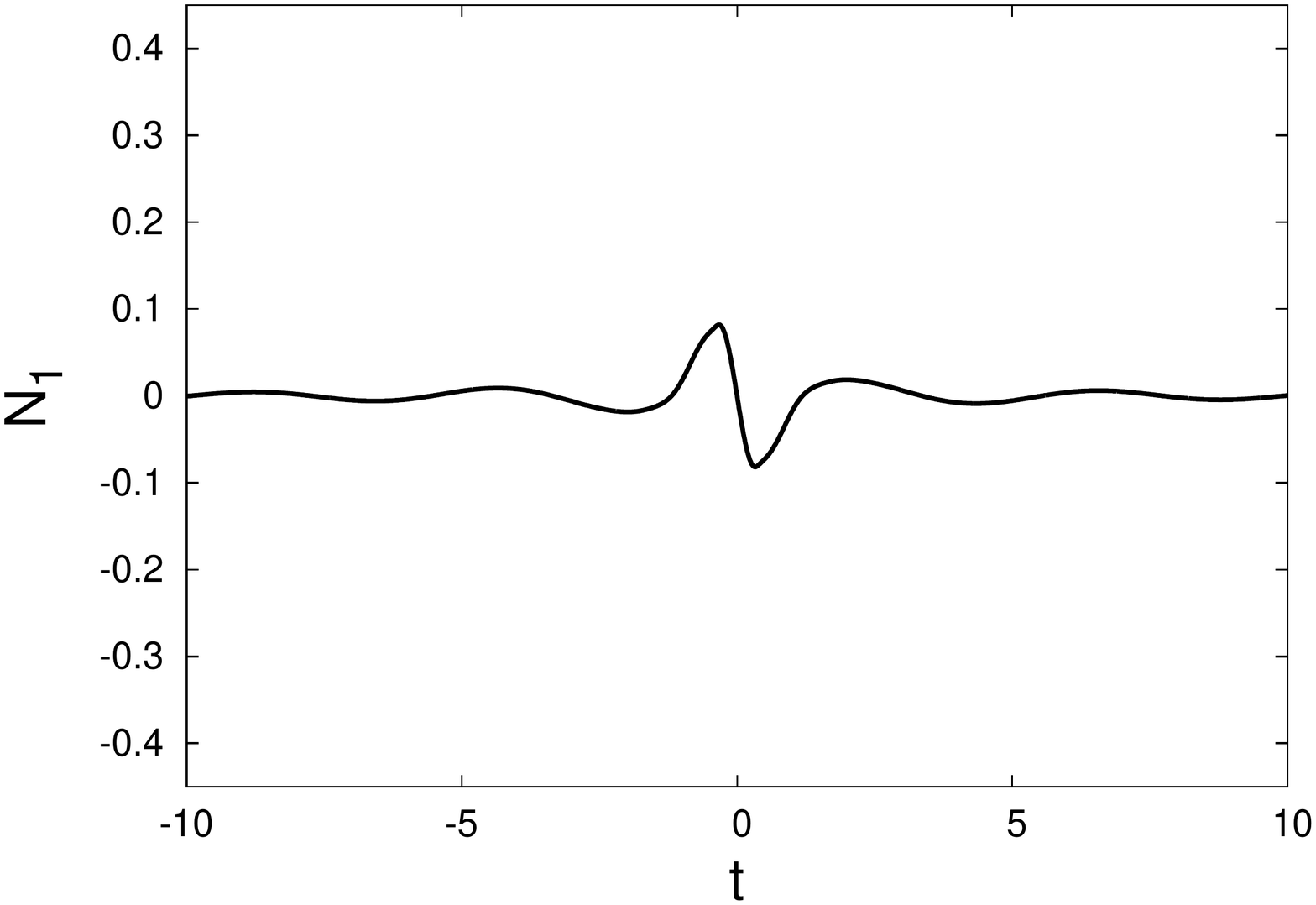}}
\\
\scalebox{0.3}{\includegraphics[bb = 0 50 654 770]{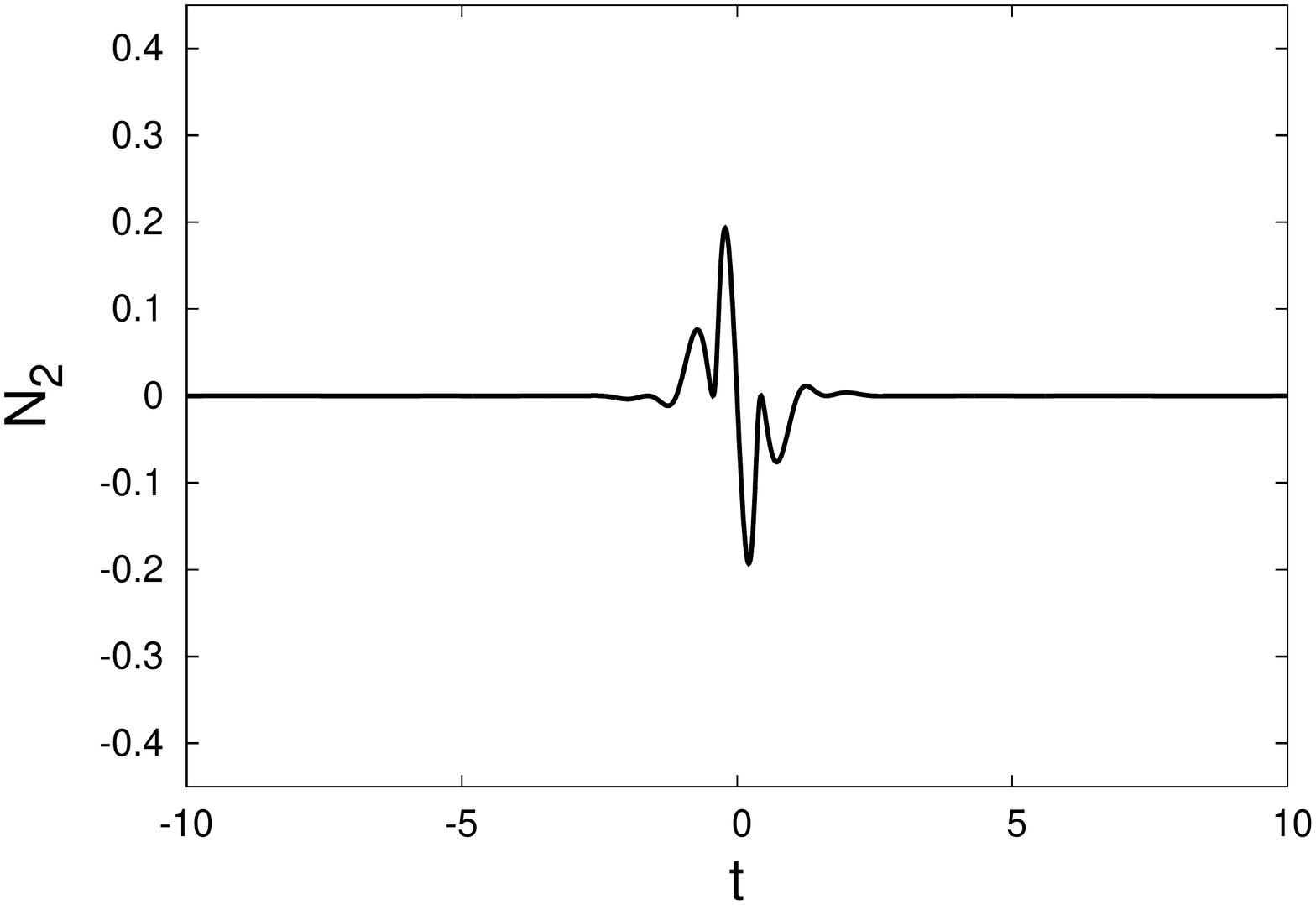}}
\hfill
\scalebox{0.3}{\includegraphics[bb = -100 50 754 770]{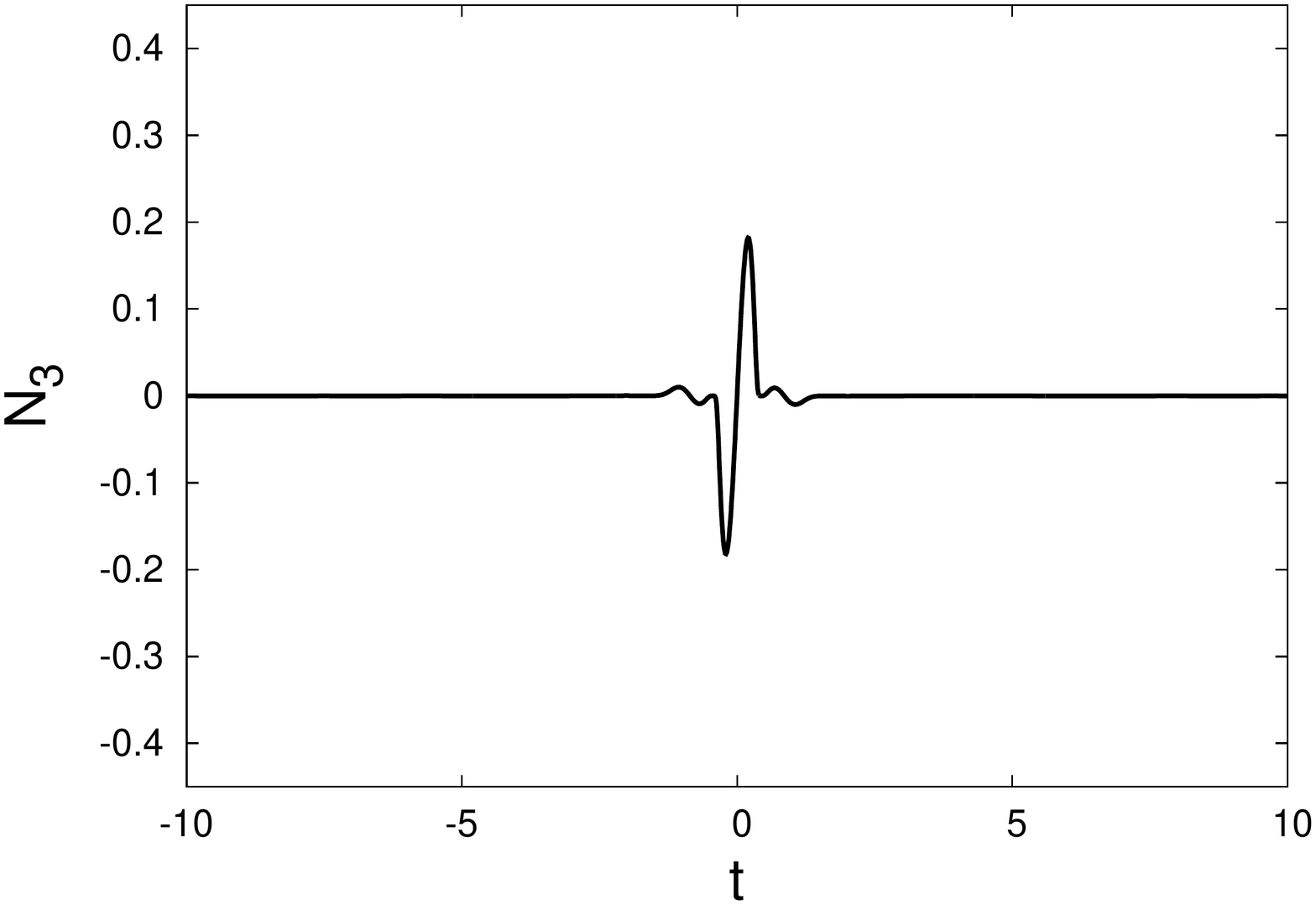}}
\caption{Comparison of the four correction terms from eq. (\ref{correction_terms}), clockwise starting at the 
upper left figure: $L_1$, $N_1$, $N_2$, and $N_3$, for a Sakovich soliton as initial condition at $x=0$.}
\label{fig:figure2}
\end{figure}

Figure \ref{fig:figure2} shows all four terms, $L_1$, $N_1$, $N_2$, and $N_3$ for the Sakovich-soliton with $m=0.35$ and, clearly, the nonlocal dispersive term is much stronger than the nonlinear terms.
This numerical observation leads to a  {\em{truncated}} higher-order SPE, that only involves the linear
term and, therefore, is much easier to compute and to analyze:

\begin{equation}
-2A_{\varkappa \phi} = \chi_0 A+ \chi_3 (A^3)_{\phi \phi} - \epsilon^2 \frac{\chi_0^2}{4} \int_{-\infty}^{\phi} d\tilde\phi \int_{-\infty}^{\tilde\phi} A(\phi_{'})d\phi^{'} \,.
\end{equation}

\begin{figure}[htb]
\centering
\scalebox{0.4}{\includegraphics[bb = 150 50 554 570]{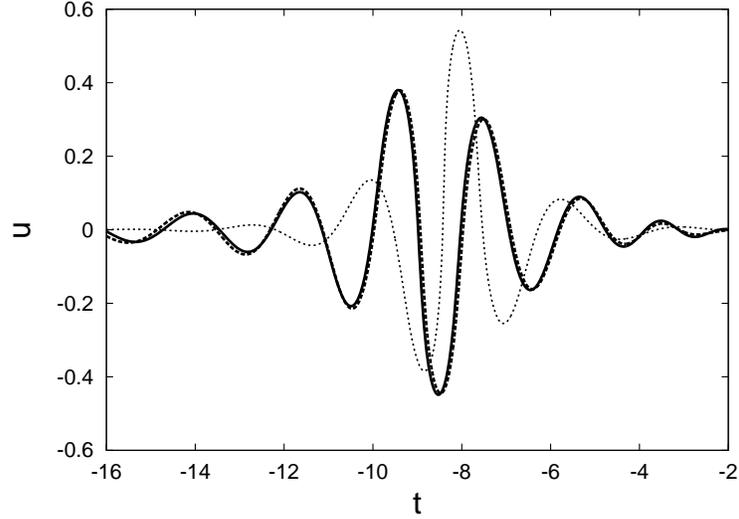}}
\hfill
\caption{Snapshot of the evolution of the Sakovich soliton in the HSPE (solid line), the truncated HSPE (dashed line) and the SPE (dotted line). }
\label{fig:figure3}
\end{figure}

Figure \ref{fig:figure3} shows the numerical solutions of the HSPE and the truncated HSPE in comparison to the numerical solution of the SPE, again with a Sakovich soliton being the initial condition. This figure shows that the difference in the performance of truncated HSPE and HSPE is very small and the truncated HSPE is still a better approximation to the full nonlinear wave equation than the SPE. This implies that, for practical purposes in some cases, the truncated HSPE might provide sufficient additional accuracy.

\subsection{Multi-soliton propagation}

Multi-soliton solutions of the SPE have been derived \cite{matsuno:2007} 
through a systematic procedure from breather solutions of the sine-Gordon equation. The parametric multi-soliton solution can be expressed in the compact form as
\begin{eqnarray}\label{spe_soln_Matsuno_MultiSoln}
u(y,t)=2i\left(\ln\frac{f^{'}}{f}\right)_t\,, \qquad x(y,t)=y-2\left(\ln f^{'}f\right)_t+d 
\end{eqnarray}
with
\begin{eqnarray}\label{spe_soln_Matsuno_MultiSoln_With}
f= \sum_{\mu=0,1} \mathrm{exp} [ \sum_{j=1}^N \mu_j \left( \chi_j+\frac{\pi}{2}i \right)+ \sum_{1 \leq j \leq N} \mu_j \mu_k \gamma_{jk} ] \\ \nonumber
f^{'}= \sum_{\mu=0,1} \mathrm{exp} [ \sum_{j=1}^N \mu_j \left( \chi_j-\frac{\pi}{2}i \right)+ \sum_{1 \leq j \leq N} \mu_j \mu_k \gamma_{jk} ] \\ \nonumber
\xi_j=p_jy+\frac{1}{p_j}t+\xi_{j0}, \qquad (j=1,2,...,M)\\ \nonumber
e^{\gamma_{jk}}=(\frac{p_j-p_k}{p_j+p_k})^2, \qquad (j,k=1,2,...,M; j \neq k) \\ \nonumber
p_{2j-1}=p_{2j}^{*} \equiv a_j+ib_j, \quad a_j>0, \quad b_j>0, \quad (j=1,2,...,M) \\ \nonumber
\xi_{2j-1,0}=\xi_{2j,0}^{*}\equiv \lambda_j+i\mu_j,   \quad (j=1,2,...,M) \\ \nonumber
\theta_j=a_j(y+c_jt)+\lambda_j,  \quad (j=1,2,...,M) \\ \nonumber
\chi_j=b_j(y-c_jt)+\mu_j,  \quad (j=1,2,...,M) \\ \nonumber
c_j=\frac{1}{a_j^2+b_j^2},  \quad (j=1,2,...,M) \,,
\end{eqnarray}
where $p_j$ and $\xi_{j0}$ are arbitrary parameters such that $p_j\neq\pm p_k$ for $j\neq k$, $i=\sqrt{-1}$, $N$ is an arbitrary positive integer, and $M=N/2$ is the number representing the multi-soliton solutions (one soliton, two solitons, etc.). If $N=4$ and $M=2$ are chosen, one can generate the two-soliton solution. The condition for a single-valued nonsingular multi-breather solution is
\begin{equation}\label{MultiSolitonCondition}
0<\sum_{j=1}^M\frac{a_j}{b_j}<\sqrt{2}-1\,.
\end{equation}  

%
\begin{figure}[htb]
\centering
\scalebox{0.4}{\includegraphics[bb = 150 50 554 570]{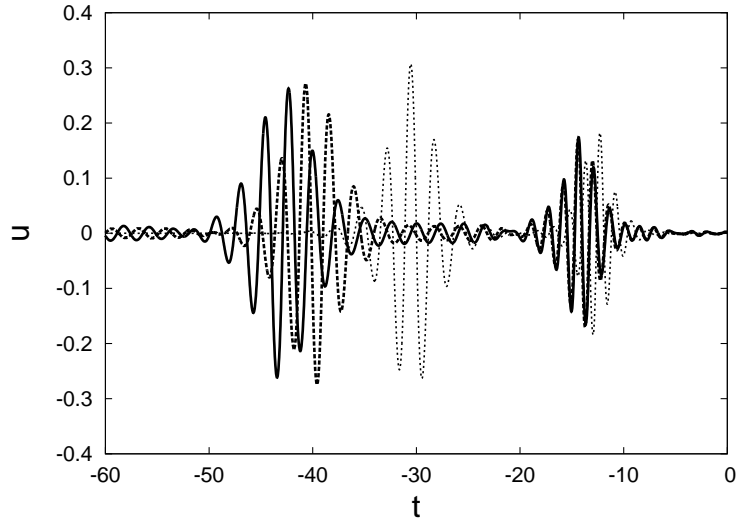}}
\hfill
\caption{Snapshot of the evolution of the SPE two-soliton solution (\ref{spe_soln_Matsuno_MultiSoln}) in the nonlinear wave equation (\ref{maxwell_1d}) (solid line), the HSPE (\ref{hspe}) (dashed line), and the SPE (dotted line).}
\label{fig:figure4}
\end{figure}

In order to investigate collisions of SPE-solitons in the HSPE and the nonlinear wave equation, we use a two-soliton solution as the initial condition. The parameters are chosen $a_1=0.1$, $b_1=0.5$, $a_2=0.16$, $b_2=0.8$, $d=0$, $\lambda_1=10$, $\lambda_2=0$, $\mu_1=0$, $\mu_2=0$. Here, the propagation distance needs to be much longer in order to capture the entire collision. Therefore,
we choose $\epsilon = 0.2$ and a propagation distance $x = 375$. At such large
distances, higher-order terms start to influence the propagation as it can be seen in figure \ref{fig:figure4}. Clearly, the HSPE approximates
the solution of the nonlinear wave equation much better than the SPE. However, it seems to be necessary to incorporate 
even higher order terms in the approximation derived in section \ref{sec:derivation}. Extending the results of section \ref{sec:derivation}, it is easy to see that the next linear term in the 
expansion is given by an integral operator involving four integrations of the form
\begin{equation}
\epsilon^4M(A) = \epsilon^4\frac{\chi_0^3}{8}\int_{-\infty}^{\phi}d\phi_1 \int_{-\infty}^{\phi_1}d\phi_2\int_{-\infty}^{\phi_2}d\phi_3\int_{-\infty}^{\phi_4}d\phi_4
A(\varkappa,\phi_4) \label{lin_term}
\end{equation}
yielding the equation
\begin{equation}\label{hspe}
-2A_{\varkappa \phi} = \chi_0 A+ \chi_3 (A^3)_{\phi \phi} + \epsilon^2 N(A) + \epsilon^4 M(A)\,.
\end{equation}

In a heuristic way, this additional term can be understood in the following way: Consider again equation (\ref{B_1}), 
we see that higher-order contributions will arise from  $-B_{xx}$ in (\ref{B_1}), in particular a term
$-2A_{x_1x_3}$ which will be renormalized by introducing a scale $x_5$.  From equation (\ref{hspe}),  we find that $2A_{x_3\phi} = A_{x_1x_1}$. Therefore, keeping only linear terms, we obtain
%
%
%
%
%
\begin{equation}
A_{x_3} \approx \frac{\chi_0^2}{8} \int_{-\infty}^{\phi}d\phi_1 \int_{-\infty}^{\phi_1}d\phi_2\int_{-\infty}^{\phi_2}d\phi_3 A
\end{equation}
and hence,
\begin{equation}
A_{x_3x_1} \approx -\frac{\chi_0^3}{16} \int_{-\infty}^{\phi}d\phi_1 \int_{-\infty}^{\phi_1}d\phi_2\int_{-\infty}^{\phi_2}d\phi_3 \int_{-\infty}^{\phi_4}d\phi_4 A
\end{equation}
yielding the addtional linear term in the HSPE (\ref{hspe}).

\begin{figure}[htb]
\centering
\scalebox{0.4}{\includegraphics[bb = 150 50 554 570]{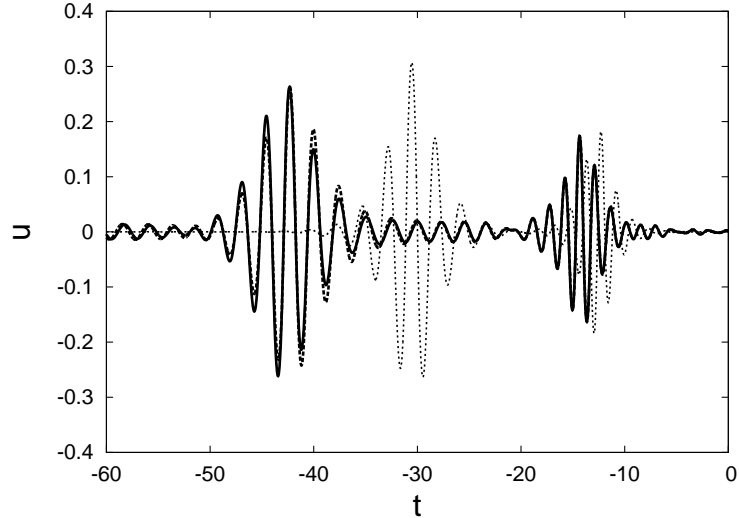}}
\hfill
\caption{Snapshot of the evolution of the SPE two-soliton solution (\ref{spe_soln_Matsuno_MultiSoln}) in the nonlinear wave equation (\ref{maxwell_1d}) (solid line), the HSPE (\ref{hspe}) with the additional linear term (\ref{lin_term}) (dashed line) and the SPE (dotted line).}
\label{fig:figure5}
\end{figure}

Indeed, figure \ref{fig:figure5} shows that there is a considerable improvement for the evolution of two-soliton solution
when incorporating this additional linear term on the r.h.s. of eq. (\ref{hspe}). Without showing the corresponding figure,
we also note that, as expected, there is improvement for the approximation of the one-soliton solution.

%

\section{Conclusion}
Using the renormalization group method, we derived higher-order correction terms to the 
short-pulse equation. We showed numerically that the incorporation of these terms can 
considerably improve the accuracy of the SPE approximation when describing solutions to 
Maxwell's equations. This was shown using one- and two-soliton SPE solutions as initial conditions. Even the incorporation of one additional linear term in the SPE approximation can lead to considerable improvement, and the extension to higher-order terms is straight-forward. A precise study of the effect of the higher-order terms on the solitary waves is subject to future studies.

\section{Acknowledgments}
TS and LK thank David Trubatch for many useful discussions. LK and TS were partially supported by the PSC-CUNY award PSC-REG-41-991.
TS was partially supported by NSF through the grants DMS-0807396 and CNS-0855217.

\clearpage


\section*{References}

\begin{thebibliography}{10}
\expandafter\ifx\csname url\endcsname\relax
  \def\url#1{\texttt{#1}}\fi
\expandafter\ifx\csname urlprefix\endcsname\relax\def\urlprefix{URL }\fi

\bibitem{agrawal:2007}
G.~P. Agrawal, Nonlinear Fiber Optics, Academic Press, 2007.

\bibitem{brunelli:2005}
J.~C. Brunelli, The short pulse hierarchy, J. Math. Phys. 46 (2005) 123507.

\bibitem{brunelli:2006}
J.~C. Brunelli, The bi-hamiltonian structure of the short pulse equation, Phys.
  Lett. A 353 (2006) 475--478.

\bibitem{chen:1994}
L.~Chen, N.~Goldenfeld, Y.~Oono, Renormalization group theory for global
  asymptotic analysis, Phys. Rev. Lett. 73 (1994) 1311--15.

\bibitem{chen:1996}
L.~Chen, N.~Goldenfeld, Y.~Oono, The renormalization group and singular
  perturbations: Multiplescales, boundary layers and reductive perturbation
  theory, Phys. Rev. E 54 (1996) 379--94.

\bibitem{chung-jones-etal:2005}
Y.~Chung, C.~Jones, T.~Sch{\"a}fer, C.~E. Wayne, Ultra-short pulses in linear
  and nonlinear media, Nonlinearity 18 (2005) 1351--1374.

\bibitem{costanzino:2006}
N.~Costanzino, Ph.D. Thesis: Existence and Stability of Nonlinear Wave
  Structures in One and Several Space Dimensions, Division of Applied
  Mathematics at Brown University, Providence, Rhode Island, 2006.

\bibitem{costanzino-manukian-jones:2008}
N.~Costanzino, V.~Manukian, C.~Jones, {Solitary waves of the regularized short
  pulse and Ostrovsky equations}, arXiv.org nlin (2008) 0809.3294.

\bibitem{erbas:2008}
K.~Erbas, Master Thesis: Some Properties and Conserved Quantities of The Short
  Pulse Equation, The Graduate School of Natural Sciences and Applied Sciences
  of Middle East Technical University, Ankara, Turkey, 2008.

\bibitem{essiambre-agrawal:1997}
R.~J. Essiambre, G.~P. Agrawal, Timing jitter of ultrashort solitons in
  high-speed communication systems. i. general formulation and application to
  dispersion-decreasing fibers, J. Opt. Soc. Am. B 14~(2) (1997) 314--322.

\bibitem{fork-cruz-becker-shank:1987}
R.~L. Fork, C.~H.~B. Cruz, P.~C. Becker, C.~V. Shank, Compression of optical
  pulses to six femtoseconds by using cubic phase compensation, Opt. Lett. 12
  (1987) 483--485.

\bibitem{kassam-trefethen:2005}
R.~L. Fork, C.~H.~B. Cruz, P.~C. Becker, C.~V. Shank, Aly-khan kassam and
  llodyd n. trefethen, SIAM J. Sci. Comput. 26~(4) (2005) 1214--1233.

\bibitem{Levent2011}
L.~Kurt, Modeling of ultra-short solitons in deterministic and stochastic
  nonlinear cubic media, {Ph.D.} thesis, The Graduate Center and University
  Center, The City University of New York (2011).

\bibitem{manukian-costanzino-etal:2009}
V.~Manukian, N.~Costanzino, C.~K.~R.~T. Jones, B.~Sandstede, {Existence of
  Multi-Pulses of the Regularized Short-Pulse and Ostrovsky Equations}, Journal
  of Dynamics and Differential Equations 21 (2009) 607--622.

\bibitem{matsuno:2007}
Y.~Matsuno, Multiloop soliton and multibreather solutions of the short pulse
  model equation, Journal of the Physical Society of Japan 76~(8) (2007)
  084003--084008.

\bibitem{parkes:2008}
E.~Parkes, Some periodic and solitary travelling-wave solutions of the
  short-pulse equation, Chaos, Solitons \& Fractals 38~(1) (2008) 154--159.

\bibitem{pietrzyk-kanatt-bandelow:2008}
M.~Pietrzyk, I.~Kanatt\u{s}ikov, U.~Bandelow, On the propagation of vector
  ultra-short pulses, Journal of Nonlinear Mathematical Physics 15~(2) (2008)
  162--170.

\bibitem{rothenberg:1992}
J.~E. Rothenberg, Space-time focusing: breakdown of the slowly varying envelope
  approximation in the self-focusing of femtosecond pulses, Opt. Lett. 17
  (1992) 1340--1342.

\bibitem{sakovich-sakovich:2005}
A.~Sakovich, S.~Sakovich, The short pulse equation is integrable, J. Phys. Soc.
  Jpn. 74 (2005) 239--241.

\bibitem{sakovich-sakovich:2006}
A.~Sakovich, S.~Sakovich, Solitary wave solutions of the short pulse equation,
  J. Phys. A: Math. Gen. 39 (2006) L361--L367.

\bibitem{sakovich-sakovich:2007}
A.~Sakovich, S.~Sakovich, On transformations of the rabelo equations, SIGMA 3
  (2007) 086.

\bibitem{sakovich:2008}
S.~Sakovich, Integrability of the vector short pulse equation, Journal of the
  Physical Society of Japan 77 (2008) 123001.

\bibitem{schaefer-wayne:2004}
T.~Sch{\"a}fer, C.~E. Wayne, Propagation of ultra-short optical pulses in cubic
  nonlinear media, Physica D 196 (2004) 90--105.

\bibitem{sulem-sulem:1999}
C.~Sulem, P.~Sulem, The Nonlinear Schr{\"o}dinger Equation Self-Focusing and
  Wave Collapse, Springer-Verlag, 1999.

\bibitem{weiner:1995}
A.~M. Weiner, Femtosecond Optical Pulse Shaping and Processing, Pergamon, 1995.

\end{thebibliography}

\end{document}